\documentclass[preprints,article,accept,moreauthors,pdftex]{Definitions/mdpi} 

\firstpage{1} 
\pubvolume{xx}
\issuenum{1}
\articlenumber{5}
\pubyear{2020}
\copyrightyear{2020}
\history{Received: date; Accepted: date; Published: date}

\Title{A Key-Agreement Protocol Based on Static Parameters and Hash Functions}

\Author{Behrooz Khadem $^{1}$*\orcidA{0000-0002-7651-7415}, Amin Masoumi $^{1}$ and M. S. Farash $^{2}$} 

\AuthorNames{Behrooz Khadem, Amin Masoumi and M. S. Farash}

\address{%
$^{1}$ \quad Information Technology and Communication Faculty, Imam Hossein University, Tehran, Iran; Bkhadem@ihu.ac.ir\\
$^{2}$ \quad Faculty of Mathematical Sciences and Computer, Kharazmi University, P. O. Box 15719-14911, Tehran, Iran; sabzinejad@khu.ac.ir}

\corres{Behrooz Khadem: Bkhadem@ihu.ac.ir; }

\abstract{Wireless Body Sensor Network (WBSN) is a developing technology with constraints in energy consumption, coverage radius, communication reliability. Also, communications between nodes contain very sensitive personal information in which sometimes due to the presence of hostile environments, there are a wide range of security risks. As such, designing authenticated key agreement (AKA) protocols is an important challenge in these networks. Recently, Li et al. proposed a lightweight scheme using the hash and XOR functions which is much more efficient compared with similar schemes based on elliptic curve. However, the investigations revealed that the claim concerning the unlinkability between the sessions of a sensor node is NOT true. The present paper considers the security issues of the scheme proposed by Li et al. and some of its new extensions in order to propose a new AKA scheme with anonymity and unlinkability of the sensor node sessions. The results of theoretical analysis compared with similar schemes indicate that the proposed scheme reduces average energy consumption and average computation time by 61 percent while reduces the average communication cost by 41 percent. Further, it has been shown by formal and informal analysis that, Besides the two anonymity and unlinkability features, the other main features of the security in the proposed scheme are comparable and similar to the recent similar schemes.}


\begin{document}


\section{Introduction}
\label{intro}

The infrastructural weaknesses, uncontrollable environments and the nature of wireless communications have caused wireless body sensor network (WBSN) to face many information security challenges. The lightweight authentication and key-agreement mechanism has been addressed by many studies in the last decade because guaranteeing a secure link in the session initiation protocol requires a secure mutual authenticated key-agreement (AKA) to secure next communications.

WBSN offers many advantages that makes it attractive for the researcher and industry \cite{Hu11}. As such, it has been proposed in order to facilitate an effective paradigm for e-health and telemedicine requirements \cite{LBMBD11}. Using miniature sensor nodes on the body (wearable) or inside the body (implantable) allows for dynamic monitoring of the physiological information such as blood pressure, heart rate, and body temperature without interfering with daily activities \cite{Ul12}. The main goal of this network is monitoring, which includes tracking and recording the vital and significant changes of  the health of the patients \cite{Ht14}.

After introducing the IEEE Standard 802.15.6 (WBAN) in 2012 \cite{Aa12}, as a promising wireless technology for low-power devices, numerous cryptography protocols were proposed based on the sensitivity, universality, and mobility of the network \cite{Sp12}. AKA allows the server in such networks to verify the user ID when accessing the system and attempting to generate a session key. Hence, in these networks one of the solutions for countering the threats and increasing the security is using key-agreement protocols to generate a session key between the sensor node and the hub node. Recently, it has become a challenge to provide such schemes with sensor node anonymity and session unlinkability. This is one of the primary motives behind the current paper.

Ibrahim et al. (2016) proposed a lightweight and efficient mechanism for mutual authenticated key agreement based on the anonymity and unlinkability of the sensor node \cite{Im16}. Their solution was very efficient for recourse-constrained sensors; because it only used hash and XOR computations instead of elliptic curve cryptography \cite{Hj12,IAU13,Sj15,Yk16}. Later, Li et al. (2017) improved upon the previous solution in the two-hop network \cite{Lx17}. The scheme starts via pre-deployed keys and parameters by the system administrator and after the mutual authentication of the nodes, a new session key generates. The main targets of the solution was to decrease the cost of complexity with maintaining the anonymity and unlinkability of the sensor node during the protocol execution.

Nevertheless, Chen et al. (2018) showed that the claim of Li et al. according to the unlinkability feature between the sessions was not true and thus, they proposed a new improved scheme by adding a predefined parameter and hash functions \cite{Cc18}. On the other hand, Khan et al. (2018) analyzed the scheme proposed by Li et al. and in addition to the problem found by Chen et al., pointed out the anonymity dilemma of the protocol \cite{KDM18}. Therefore, they proposed an improved scheme by adding the additional hash functions and an auxiliary parameter for mutual authentication; however, their scheme increased computational complexity. Recently, Kompara et al. (2019) proposed an improved and low-complexity robust scheme. They showed that Li’s scheme requires session unlinkability and protocol structural integrity as well as better security against threats such as jamming \cite{KIH19}.

\section{Motivations and Contributions} \label{pm}

In spite of the introducing some improved schemes, there are still limitations in WBSN communications. Therefore, the current paper attempts to propose a lightweight scheme based on the previous ones that is suitable for tow-hop or tow-tire centralized WBSN. Basically, one of the most important characteristic of sensor nodes in wireless networks is their passive mode, which is tied to protocol design. we are looking for reducing the online/live statement in a special period of time or reducing the channel rate to run an efficient key agreement protocol. Therefore, when we discuss about the nature of the sensors, it means that the sensor node does not require to be always active/present/online to create a new session key. The contributions of this paper are as follows:

\begin{enumerate}[leftmargin=*,labelsep=4.9mm]
	\item Optimizing protocol structure via refreshing the auxiliary authentication parameters at the beginning of each session without the requirement of updating the pre-deployed parameters to decrease waiting-time;
	\item Unlinkability between the subsequent sessions of a sensor node  according to the anonymity;
	\item Validating the security of the new scheme via official (formal) an non-official proofing;
	\item Reducing the computational complexity, energy consumption, and transmission bandwidth of the network.
\end{enumerate}

This paper is organized as follows: Section \ref{pre} discusses some basic concepts. Section \ref{propos} presents the proposed scheme. Section \ref{analys} analyzes the conventional security attributes. Section \ref{formal} presents a formal verification of the scheme using Scyther tool. In Section \ref{effic} by evaluating the proposed scheme and comparing with previous one, it has been shown how it reduces the computational cost, communication costs, and energy consumption. Finally, the conclusion is given in Section \ref{conc}.

\section{Prerequistic Concepts} \label{pre}

\begin{figure}[h]
	\centering
	\includegraphics[width=0.7\linewidth]{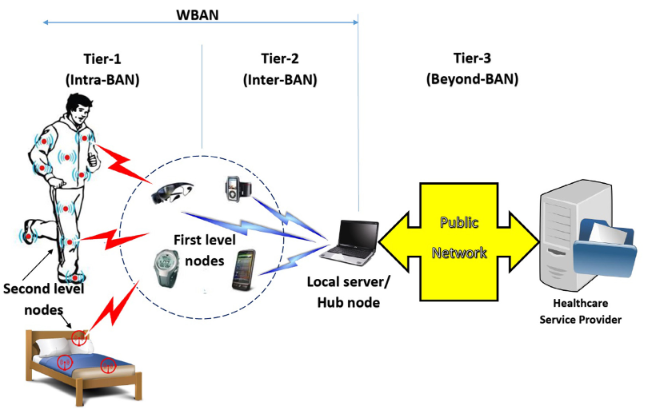}
	\caption{Three-tire network architecture}
	\label{fig1}
\end{figure}

Figure \ref{fig1} shows how the general WBSN architecture can be classified into three-tire \cite{Cm11,NJB16}.

\begin{enumerate}[leftmargin=*,labelsep=4.9mm]
	\item Tire-1 (Intra-BAN): It consists of physically sensor nodes and personal servers (i.e. a smartphone or smart watch).
	\item Tire-2 (Inter-BAN): A layer which adds an access point to the network, in a way that the personal server connects and route to the other wireless networks.
	\item Tire-3 (Beyond-BAN): consists of other wireless or public area networks that transmits the collected data to caregiver terminal database (CT).
\end{enumerate}

\begin{figure}[h]
	\centering
	\includegraphics[width=0.5\linewidth]{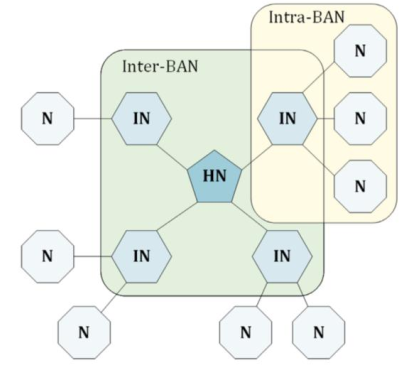}
	\caption{Two-tire WBSN model}
	\label{fig2}
\end{figure}

The centralized two-hop or two-tier network proposed by Li consists of three node types (Figure \ref{fig2}):
\begin{enumerate}[leftmargin=*,labelsep=4.9mm]
	\item SN or N: the second level node (i.e. the sensor node implanted in body). These nodes usually have limited computational and communicational power due to resource constraints.
	\item FN or IN: the first-level node that can also be referred to the intermediate or coordinator node (i.e. smartphone or smartwatch) which has a higher computational and communicational power and storage in compare with the SN.
	\item HN: the hub node or a local server (e.g. personal computer). This node has higher computational power and resources than others. It is assumed that, the hub nodes are always in-range of intermediate nodes.
\end{enumerate}

The HN and IN are the intermediate parts of the body sensor network and occupy the second level of the network. The link between the IN and N is the internal part and is assumed as the first network level. Note that HN is outside the coverage of SN, therefore after monitoring and collecting the vitals, the SNs in the internal part send the vitals of the patient to the HN via the IN immediately.

The HN retains the authentication data of the patient and processes them after collecting the vital signals from the internal part. After that, it transmits them to medical service servers based on the priority of critical information.

In Li et al. scheme, the security prerequisites follow the Dolev-Yao threat model \cite{DY83}. That is, the parties are communicating via an insecure channel. Thus, the adversary can eavesdrop on all the communication links in the network and is able to replay and modify the messages.

The HN is assumed trustable; nevertheless, the adversary may penetrate the HN database and then steal or tamper with the database information. Note that only the master key of the HN is assumed completely inaccessible to the adversary. Furthermore, any SN may be captured, that means, the adversary can extract the secret information stored in the sensor memory. Therefore, we use the threat model utilized in Li et al. scheme, too.

\section{The proposed protocol}\label{propos}

The proposed scheme consists of three phases same as previous schemes \cite{Im16,Lx17,Cc18,KDM18,KIH19}. The initialization phase and registration phase are performed by the system administrator in a secure environment. The authentication phase is performed in a public channel using a new idea.

The SN is assumed as the second level node and contacts the HN via the intermediate node (IN). If the SN be assumed as the first level node, the scheme can be adapted to direct communication with the HN. Table \ref{tab1} shows the symbols and abbreviations used in this paper.

\begin{table}[h]
	\centering
	\caption{Symbols and abbreviations}
	\begin{tabular}{cc}
	\toprule
	\textbf{Symbol} & \textbf{Description} \\ 
			\midrule 
		SA & System Administrator \\ 
		\hline 
		N & First Level node-Sensor Node \\ 
		\hline 
		IN & Second Level Node/Intermediate Node \\ 
		\hline 
		HN & Hub Node  \\ 
		\hline 
		$ id_{N} $ & Identity of SN \\ 
		\hline 
		$ id'_{IN} $ & Short identity of First level sensor node \\ 
		\hline 
		$ K_{HN} $ & Secret key of Hub Node \\ 
		\hline 
		$ K_{N} $ & Secret key-parameter of Sensor Node \\ 
		\hline 
		$ r_{N} , r_{H} $ & Temporary secret parameters as a nonce \\ 
		\hline 
		$ a_{N} , b_{N} $ & Authentication parameters  \\ 
		\hline 
		$ m_{1} , m_{2} , m_{3} , m_{4} $ & Auxiliary parameters required for authentication  \\ 
		\hline 
		$ t_{N} , t_{H} $ & Timestamp generated by SN-HN \\ 
		\hline 
		$ t_{c} , t*_{c} $ & Constant message received time \\ 
		\hline 
		$ k_{S} $ & Session Key to be agreed upon \\ 
		\hline 
		h(.) & Collision-resistant One-way cryptographic hash function \\ 
		\hline 
		$ \parallel $ & Concatenation operation \\ 
		\hline 
		$ \bigoplus $ & Bitwise XOR operation \\ 
		\hline 
		$ X \rightarrow Y:M $ & Entity X transmit the message M to entity Y via a public channel \\ 
		\bottomrule
	\end{tabular}
	\label{tab1}
\end{table} 

Also, assume  h:$ \mathit{ \left\lbrace 0,1 \right\rbrace }^{*} \longrightarrow \left\lbrace 0,1\right\rbrace ^{t} $ as a powerful hash function.

\subitem 
A.	\textit{Initialization phase}

In this phase, the system administrator, initializes the hub node \textit{HN} and assigns a master key $ K_{HN} $ to $ HN $. Then he stores $ K_{HN} $  in  the $ HN $ memory.

\subitem 
B.	\textit{Registration phase}

At this phase, first the system administrator registers the sensor node by devoting a unique ID($ id_{N} $ ) and a secret parameter ($ K_{N} $) generated for the  in \textsl{SN} the network. Then computes the Eq. (1) and (2) and stores the tuple $ < id_{N},a_{N},b_{N} > $ in the \textsl{SN} memory.

\begin{equation} \label{eq1}
a_{N}=h(id_{N} \parallel K_{N})
\end{equation}
\begin{equation} \label{eq2}
b_{N}=K_{HN} \oplus K_{N} \oplus id_{N}
\end{equation}

On the other hand, a table containing all $ K_{N} $ values assigned to the nodes as well as all  values computed by the corresponding $ K_{N} $ store in the $ HN $ memory. If there is an $ \textit{IN} $ in the network, \textsl{SA} picks a short unique ID($ id_{IN} $) and stores in $ \textit{IN} $ memory. The $ id_{IN} $ also store in the $ HN $ memory. Note that $ K_{HN} $ from the previous phase is already stored in the $ HN $ and therefore, the tuple  $ < id_{IN},K_{HN},T(K_{Ni} ,a_{Ni} ) > $ stores in the $ HN $ memory.

\subitem 
C.	\textit{Authentication and session key agreement phase}

\begin{figure}[t]
	\centering
	\includegraphics[width=1.1\linewidth]{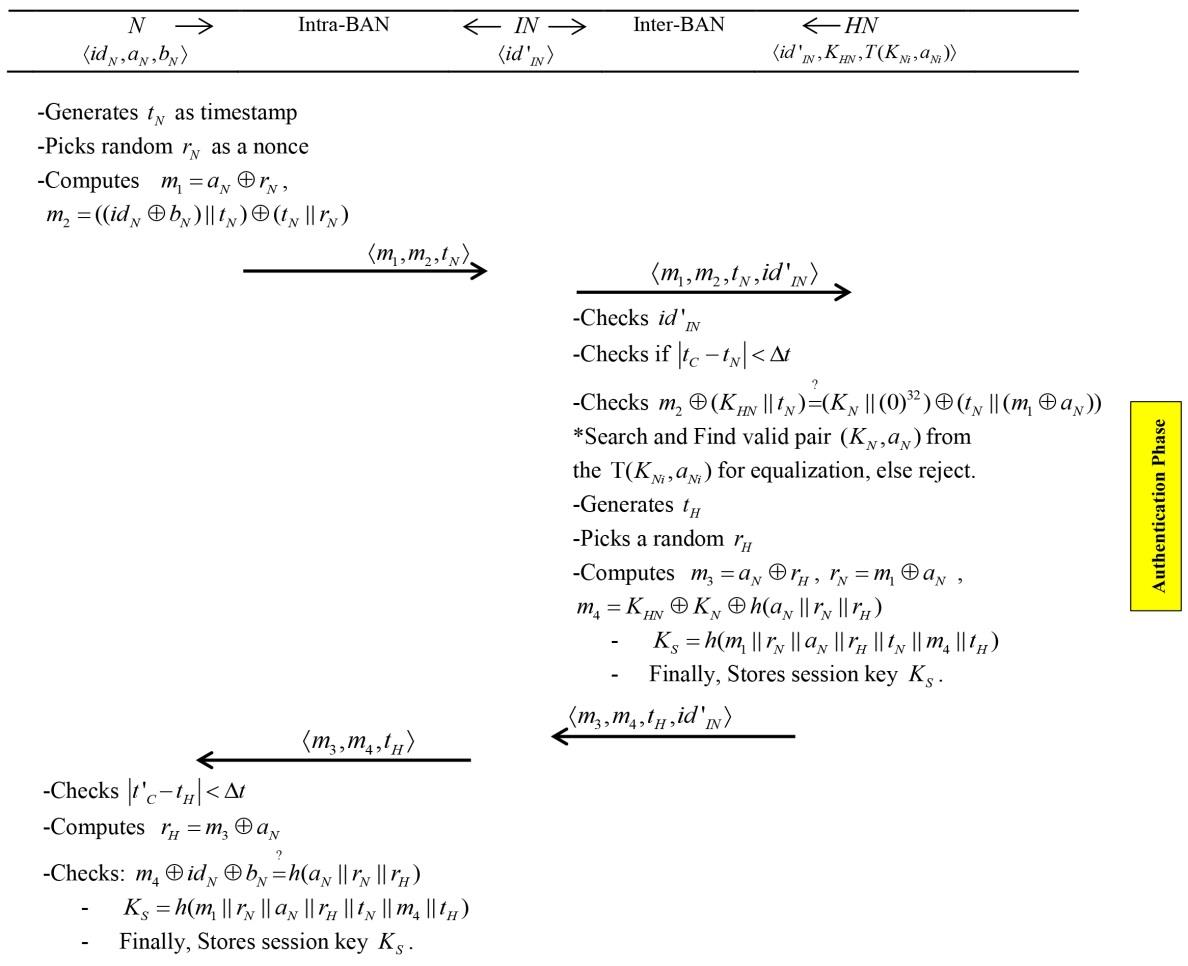}
	\caption{Authentication and key-agreement phase of the proposed scheme}
	\label{fig3}
\end{figure}

In the authentication phase, \textsl{SN} attempts for the agreement on the session key ($ K_{S} $) as an anonymously mutual authentication with the $ HN $ (Figure \ref{fig3}).

Step 1: $ N \rightarrow IN:<m_{1} , m_{2} , t_{N}> $ . The \textsl{SN} generates a timestamp $ t_{N} $ and a temporary secret parameter $ r_{N} $ as a nonce. Then it computes authentication parameters (Eq. 3 and 4) and sends the tuple $ <m_{1} , m_{2} , t_{N}> $ to the $ IN $.

\begin{equation} \label{eq3}
m_{1}=a_{N} \oplus r_{N})
\end{equation}
\begin{equation} \label{eq4}
m_{2}=((id_{N} \oplus b_{N}) \parallel t_{N}) \oplus (t{N} \parallel r_{N})
\end{equation}

Step 2: $ IN \rightarrow HN:<m_{1} , m_{2} , t_{N} , id_{IN}> $ . The $ IN $ adds its own ID ($ id_{IN} $ ) to the received message without changing the message. Then transmits the tuple $ <m_{1} , m_{2} , t_{N} , id_{IN}> $ to the $ HN $.

Step 3: $ HN \rightarrow IN:<m_{3} , m_{4} , t_{H} , id_{IN}> $. First the $ HN $ checks and verifies $ id_{IN} $ based on the value it has stored in its memory. Then it verifies the timestamp $ t_{N} $ via the default validation (Eq. 5), where $ t_{C} $ denotes the reception time of the message and $ \Delta t $ denotes the maximum transmission delay. If authentication fails, the protocol execution halts. The condition for security is that $ \Delta t $ is smaller than the minimum attack time $ t_{min}(\Delta t < t) $.

\begin{equation} \label{eq5}
|t_{C} - t_{N}|<\Delta t
\end{equation}

After verifying the received data, $ HN $ has to check and compute Eq. (6) for authenticating the \textsl{SN}.

\begin{equation} \label{eq}
m_{2} \oplus (k_{HN} \parallel t_{N}) = (K_{N} \oplus (0)^{32})\oplus (t_{N} \oplus (a_{N} \parallel m_{1}))
\end{equation}

Therefore, $ HN $ has to search in the stored table $ T(K_{Ni},a_{Ni}) $ and check $ (K_{N},a_{N}) $ pairs in the equation.(7). If the $ (K_{N},a_{N}) $ pair to satisfy the equation isn’t found, authentication fails and the   rejects the protocol execution.

After verification and successful authentication, the timestamp $ t_{H} $ is generated and the temporary secret parameter $ r_{H} $  will be picked by the $ HN $. Then Eq. (7) and authentication auxiliary parameters Eq. (8) and (9) will be computed for mutual authentication.

\begin{equation} \label{eq7}
r_{N} = m_{1} \oplus a{N} 
\end{equation}
\begin{equation} \label{eq8}
m_{3}= a+{N} \oplus r_{H}
\end{equation}
\begin{equation} \label{eq9}
m_{4}= K_{HN} \oplus K_{N} \oplus h(a_{N} \parallel r_{N} \parallel r_{H})
\end{equation}

Then, the new session key (Eq. 10) will be generated and stored. After that, the tuple $ <m_{3},m_{4},t_{H},id_{IN}> $ will be transmitted to the $ IN $.

\begin{equation} \label{eq10}
k_{s} = h(m_{1} \parallel r_{N} \parallel a_{N} \parallel r_{H} \parallel t_{N} \parallel m{4} \parallel t_{H})
\end{equation}

Step 4: $ IN \rightarrow N:<m_{3} , m_{4} , t_{H} > $ . Here the  $ IN $ acquires its  ID ($ id_{IN} $ ) and if the $ ID $ is validated, the rest of the received message $ <m_{3} , m_{4} , t_{H} > $ will be sent to the \textsl{SN}.

Step 5: $ N $ . The  \textsl{SN} receives the tuple $ <m_{3} , m_{4} , t_{H} > $  and verifies the timestamp $ t_{H} $ Then, it computes Eq. (12) and verifies Eq. (13).

\begin{equation} \label{eq11}
| t_{C} - t_{H} | < \Delta t
\end{equation}
\begin{equation} \label{eq12}
r_{H} = m_{3} \oplus a_{N}
\end{equation}

If the Eq. (13) is not satisfied, the authentication phase fails and the \textsl{SN} retries protocol execution. Therefore, after mutual authentication, the session key (Eq. 14) will be computed and stored in its memory for use in next communications.

\begin{equation} \label{eq13}
m_{4} \oplus id_{N} \oplus b_{N} = h(a_{N} \parallel r_{N} \parallel r_{H})
\end{equation}
\begin{equation} \label{eq14}
K_{s} = h(m_{1} \parallel r_{N} \parallel a_{N} \parallel r_{H} \parallel t_{N} \parallel m_{4} \parallel t_{H})
\end{equation}
\section{Security Analysis of the Proposed Protocol} \label{analys}

This section discusses about some common WBSN attacks and the security of the proposed scheme against them.

\subsection{Anonymity and Session Unlinkability}

In the proposed scheme we show that in case of seeing all compunctions via the protocol, the adversary still cannot learn the ID of any \textsl{SN}, because $ id{N} $ is not transmitted in the public channel. Parameter $ m_{1} $ has the value $ a_{N} $ and $ id_{N} $ that are protected by the one-way resistant hash function in Eq. (1).

Further, the ID in $ m_{2} $ is combined with temporary random parameters $ r_{N} , t_{N} $, and the secret $ k_{N} $ as well as the $ k_{HN} $ secret key. In addition, because $ m_{1} $ and $ m_{2} $ values are refreshed in each session using fresh secret random parameters, the adversary cannot link between a session with another successfully completed session of the same \textsl{SN}. Unlinkability is a more significant feature than anonymity, such as the generation process of a session key in the protocol cannot lead to tracing the corresponding \textsl{SN} of the session.

\subsection{Eavesdropping Attack}

In the authentication phase, the adversary can listen to and store all transmitted data; however, according to the anonymity feature, she cannot use the transmitted messages to obtain the $ a_{N} $ and $ K_{N} $ values of the \textsl{SN} to compute $ K_{s} $.

\subsection{Impersonation, Replay and Man in the Middle Attacks}

Here the adversary has two goals: (1) deception the parties by impersonating a legal partner and (2) selecting the best strategy to correctly wild guess session key identifying against a bit string equal to the key length. Therefore, due to using a timestamp, if a temporal inconsistency detect (i.e. $ |t_{C} - t_{N} > \Delta t $ or $ |t'_{C} - t_{H}| > \Delta t $ ), the message will be rejected. In addition, the fresh random values $ r_{N} $ and $ r_{H} $ reveal any modification of the transmitted message, therefore, the adversary needs to be able to change the parameters of the transmitted messages to spoof the messages. However, for this purpose she needs to know $ id_{N} $,$ a_{N} $ and $ b_{N} $ , A task which is impossible via eavesdropping unless the adversary captures the \textsl{SN}. Nevertheless, compromising the $ K_{HN} $  is not only NOT possible, But also assumed completely beyond the reach of the adversary according to Dolev-Yao threat model and on the other hand it will be protected by the random temporary value $ K_{N} $ in Eq. (2).

\subsection{Sensor Node Capture Attack and Intruding to the IN}

Suppose the attacker physically captures the sensor node. In this case, it should be seen whether other sensor nodes in relationship with the hub node will be compromised as a result of the forgery and disclosure of that one sensor node or not. Hence, disclosure or forging of the sensor node tuple $ <id_{N},b_{N},a_{N}> $  will not provide any important information to the adversary. Because of calling $ a_{N}= h(id_{N} \parallel K_{N}) $ , he is not aware of the secret $ K_{N} $ and this secret value is protected by the one-way hash function. On the other hand, by recalling $ b_{N} \oplus id_{N} = K_{HN} \oplus K_{N} $ , the adversary has no information about secret values $ K_{HN} $ and $ K_{N} $ , so the other sensor nodes will no longer be affected by this capture node attack.
The adversary is also capable to steal the \textit{IN} (e.g. the smartphone) and penetrate it; however, she cannot impersonate the secure communication between the \textsl{SN} and \textsl{HN}. Because there are no needed long-term parameters stored in \textit{IN} memory for the adversary's calculations, and the only parameter stored in its memory is the 16-bits ID, which is used only for authentication by the \textsl{HN} and itself.

\subsection{Jamming and De-synchronization}

An authentication scheme is vulnerable to a desynchronization attack if it requires the two parties to update their state in synchronism. Therefor an authentication attempt by an adversary is blocked, the \textsl{SN} still can retry authentication request using the previous $ K_{N} $ and $ a_{N} $. Because the \textsl{HN} stores a list of $ (K_{Ni} , a_{Ni}) $ values in the memory, it can constantly keep up and synchronize itself with the repeated \textsl{SN} requests. Subtitle i represents the sensor node number.

\subsection{Forward and Backward Secrecy}

This security feature indicates that if the session key is revealed, the previous and next session keys are not exposed. While session keys can be obtained by capturing a node, Eq. (10) and (14) are protected by a one-way hash function as well as random, fresh and dynamic temporary parameters from the same session.

\section{Formal Verification via Scyther} \label{formal}

This section uses Scyther tool  \cite{Cc06,Vl06} to show that the proposed scheme is secure according to the formal proof. This tool uses Dolev and Yao’s intruder model as well as refining the encrypted schemes and has been designed for automatic verification, spoofing, and analyzing the security features of security protocols \cite{KIH19}.

\begin{figure}[t]
	\centering
	\includegraphics[width=1.1\linewidth]{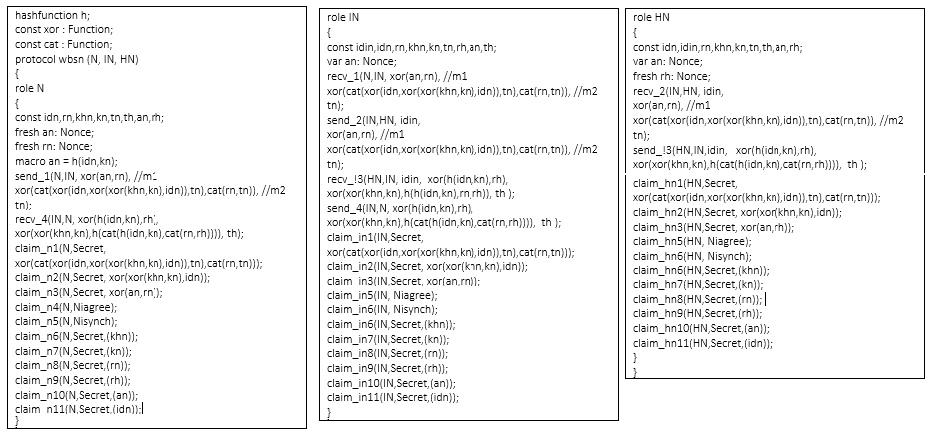}
	\caption{The SPDL code of the proposed scheme}
	\label{fig4}
\end{figure}

This tool is developed based on Python and uses the security protocol description language (SPDL). The written language can be modeled by the command line interface (CLI) and show us a window using the security verification results against different attacks. Further, if an attack be detected, it can use the graphical user interface (GUI) to present a graphical view of the detected attack \cite{KIH19}. Since the tool can be used to verify the claims defined by the user and the automatic claims generated by the tool itself, the written code following the security verifications of the authentication phase of the proposed scheme is presented at the end of this paper (Figure \ref{fig4}).

\section{Efficiency Analysis} \label{effic}
This section checks and calculates the storage space, computational costs and energy consumption of the \textsl{SN} and the \textsl{HN} on the same simulated device. Also, The communication cost (required bandwidth) has been evaluated by the length of transmitted messages. Finally, the efficiency of the proposed scheme will be compared with some previous schemes.

\subsection{Storage Requirement}
According to Section 5.1 of Li et al. paper, $ | id'_{IN}| $ is 16 bits and $ |t_{N}| = |t_{H}| = 32 $ bits. Since the hash function uses the SHA-1 algorithm, and the output $ |a_{N}| $ is 160 bits, the other parameters $ |id_{N}|,|b_{N}|,|r_{N}|,|r_{H}|,|K_{N}|,|K_{HN}|$ , and $ |K_{s}| $ are also assumed 160 bits. Therefore, the storage space required by the $ SN(|id_{N},a_{N},b_{N},K_{s}) $ is 640 bits while the \textsl{HN} storage ( $ |K_{Si} $ and $ |id'_{IN},K_{HN},T(K_{Ni},a_{Ni})| $) is $ 480n+16m+160 $ bits , where $ m $ is the number of the INs and  is the number of the SNs.

\subsection{Computational Time and Cost}
\begin{figure}[t]
	\centering
	\includegraphics[width=0.7\linewidth]{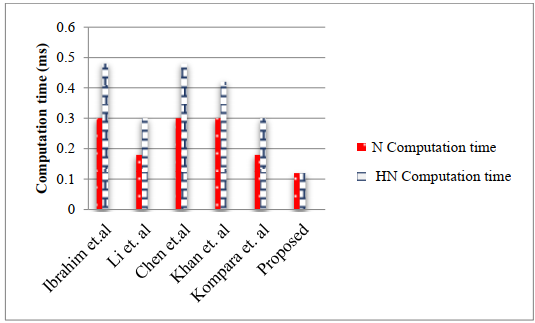}
	\caption{Computation time}
	\label{fig5}
\end{figure}

Assume $ t_{xor} $ denotes the computational time of XOR opertion and $ t_{h} $ as the computational time for 160-bit hash function. Note that $ t_{xor} $is extremely small compared with $ t_{h} $ and therefore we can assume $ t_{xor} \approx 0 $ . Accordingly, in the proposed protocol the computational costs for the \textsl{SN} and \textsl{HN} are obtained by Eq. (15) and (16). To measure the computation time of cryptographic primitives in a hardware simulated environment, a 32-bit Cortex-M3 micro-controller with 512 KB of memory and a frequency of 72 MHz was utilized \cite{Do16,Lj15}. Recalling a SHA-1 hash function takes 0.06 ms \cite{Lj15}. Hence, the computation time for two hash functions using SHA-1 is given by Eq. (17) (see Figure \ref{fig5} and Table \ref{tab2}).

\begin{equation} \label{eq15}
2t_{h} + 6t_{xor}
\end{equation}
\begin{equation} \label{eq16}
2t_{h} + (2n+5)t_{xor}
\end{equation}
\begin{equation} \label{eq14}
T_{2th}=2 \times 0.06=0.12 ms
\end{equation}

\subsection{Energy Consumption}
\begin{figure}[t]
	\centering
	\includegraphics[width=0.7\linewidth]{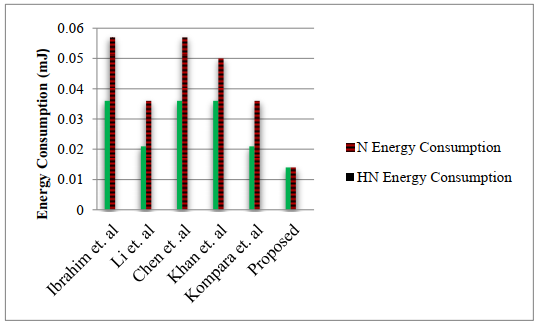}
	\caption{Energy consumption}
	\label{fig6}
\end{figure}

The consumed current by the micro-controller in Section B. at room temperature (300 K or 27C) is in active mode with a rated voltage of 3.3 V is 36 mA \cite{Do16}. Therefore, the power consumed in active state is 118.8 mW which shows a very low power consumption. Hence, according to Section B, it is possible to use the power consumption during computations for approximate estimates. Accordingly, the energy used by either the \textsl{HN} or \textsl{SN} with two hash functions are given by Eq. (18), (see Figure \ref{fig6} and table \ref{tab2}).

\begin{equation} \label{eq14}
E_{N} = E_{HN}=(0.12 \times 118.8)/1000 = 0.014 mJ
\end{equation}

\subsection{Communication Costs}
\begin{figure}[t]
	\centering
	\includegraphics[width=0.7\linewidth]{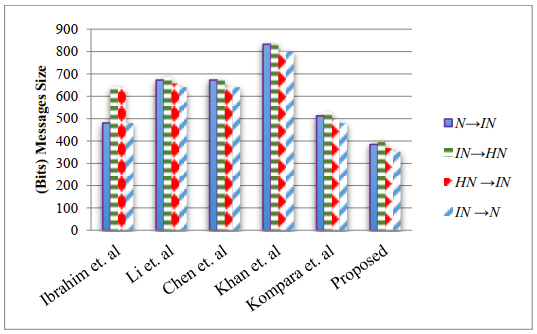}
	\caption{Communication cost}
	\label{fig7}
\end{figure}

As indicated by the authentication steps, when $ N \rightarrow IN:<m_{1},m_{2},t_{N}> $, the message is 384 bits whereas when $ IN \rightarrow HN:<m_{1},m_{2},t_{N},id'_{IN}> $ , the message is 400 bits. In the return path when $ HN \rightarrow IN:<m_{3},m_{4},t_{H},id'_{IN}> $ the message is 368 bits and in the last transmission, $ IN \rightarrow N:<m_{3},m_{4},t_{H}> $ the message is 352 bits. This results indicate that the new scheme uses less bandwidth compared with its previous counterparts, as shown in Figure \ref{fig7} and Table \ref{tab2}.

\section{Conclusion} \label{conc}

This paper presented a lightweight mutual authentication key agreement protocol for a two-tire centralized WBSN. Since body sensors are severely energy constrained, we have shown in our proposed scheme that in addition to consider the passive mode of sensor nodes as an inherent feature, the balance between security and performance can still be maintained. For this purpose we showed that for each session, the previous auxiliary parameters are refreshing before the first authentication request by the \textsl{SN} and there is no need to update the predefined parameters. Hence the auxiliary equations and hash functions are used fewer times than previous Protocols in Table \ref{tab3}, that is still retained a good balance between performance and security.

In the discussion of security, while maintaining the two criteria of anonymity and unlinkability between sessions, we investigated common vulnerabilities and with the help of Scyther security verification tool, showed that the proposed protocol achieved acceptable security. Concerning energy consumption, compared to other schemes the \textsl{SN} and \textsl{HN} respectively consumed 53 and 70 percent less energy while reducing the computation time by 52 and 70 percent, respectively. Further, the proposed scheme reduced communication costs by 41 percent compared to other previous schemes.

\section*{Acknowledgements}

Hereby, we would like to sincerely thank Mister Marko Kompara for his useful recommendations.


\begin{table}[t]
	\centering
	\caption{ Bandwidth used (communicant cost) per message (bit)}
	\begin{tabular}{ccccccc}
		\toprule 
		\textbf {Communication } & \cite{Im16} & \cite{Lx17} & \cite{Cc18}& \cite{KDM18} & \cite{KIH19} & \textbf {proposed} \\ 
		\midrule 
		$ N \rightarrow L $ & 480 & 672 & 672 & 832 & 512 & 384 \\ 
		\hline 
		$ IN \rightarrow HN $ & 640 & 688 & 672 & 832 & 528 & 400 \\ 
		\hline 
		$HN \rightarrow IN $ & 640 & 656 & 640 & 800 & 496 & 368 \\ 
		\hline 
		$ IN \rightarrow N $ & 480 & 640 & 640 & 800 & 480 & 352 \\ 
		\bottomrule
	\end{tabular} 
	\label{tab2}
\end{table}

\begin{table}[t]
	\centering
	\caption{ Comparing the efficiency features running on simulated micro-controller 32-bits cortex-m3 at 72MHz}
	\begin{tabular}{cccccc}
		\toprule 
		\textbf {Schemes } & \textbf {Node} & \textbf {Storage cost} & \textbf {Cost (cycle)} & \textbf {Time (ms)} & \textbf {Energy(mJ)} \\ 
		\midrule 
		\cite{Im16} &$ N $ & 480 bits & $ 5t_{h}+2t_{xor} \approx5t_{h}$ & 0.30 & 0.036 \\ 
		                &$ HN $ &480n+160 &$ 8t_{h}+4t_{xor}\approx8t_{h} $  & 0.48 & 0.057  \\ 
		\hline 
		\cite{Lx17} &$  N $ & 640 & $ 3t_{h}+7t_{xor}\approx3t_{h}  $ &0.18 & 0.021 \\ 
	                 &$ HN $ & 160(n+1)+16m &  $ 5t_{h}+12t_{xor}\approx5t_{h} $ & 0.3 & 0.036 \\ 
		\hline 
    	\cite{Cc18} &$ N $& 800 &  $5t_{h}+ 5t_{xor}\approx5t_{h} $ & 0.3 & 0.036  \\ 
                	&$ HN  $& 160(n+1) & $ 8t_{h}+11t_{xor}\approx8t_{h} $ & 0.48& 0.057 \\ 
    	\hline 
		\cite{KDM18} &$ N $ & 640 & $ 5t_{h}+9t_{xor}\approx5t_{h}  $&0.3& 0.036 \\ 
		              &$ HN $ &160(n+1) & $ 7t_{h}+14t_{xor}\approx7t_{h} $ & 0.42 & 0.05  \\ 
	    \hline
	   \cite{KIH19} & $ N $ & 640 &  $ 3t_{h}+6t_{xor}\approx3t_{h} $ & 0.18 & 0.021  \\ 
	                 &$ HN $ & 640n+16m+160 & $ 5t_{h}+(n+7)t_{xor}\approx5t_{h} $ & 0.3 & 0.036  \\ 
	    \hline
	   	
		\textbf{Proposed} &$ N  $& 640 & $ 2t_{h}+6t_{xor}\approx2t_{h} $ & 0.12 & 0.014  \\ 
	                &$  HN $ & 480n+16m+160 & $ 2t_{h}+(2n+5)t_{xor}\approx2t_{h} $ & 0.12 & 0.014  \\ 
		\bottomrule
	\end{tabular} 
	\label{tab3}
\end{table}

\reftitle{References}



\end{document}